\documentclass{mpe_report}
\usepackage{psfig}
\begin{document}
\title{The high energy emission from black holes}
\author{M. D. Caballero-Garc\'{\i}a \inst{1} \and J. M. Miller \inst{2} 
\and E. Kuulkers \inst{3} on behalf of a larger collaboration team}  
\institute{LAEFF-INTA, P.O. Box 50727, 28080 Madrid, Spain,
\and  
Department of Astronomy, University of Michigan, 500 Church Street, Ann Arbor, USA
\and
ESA/ESAC, Urb. Villafranca del Castillo, PO Box 50727, 28080 Madrid, Spain
}
\maketitle

\begin{abstract}
The origin of the high energy emission (X-rays and ${\gamma}$-rays) from black 
holes is still a matter of debate. We present new evidence that hard X-ray 
emission in the low/hard state may not be dominated by thermal Comptonization.
We present an alternative scenario for the origin of the high energy emission 
that is well suited to explain the high energy emission from GRO~J1655$-$40. 
\end{abstract}

In this study we focus on black hole transients (i.e. binary systems constituted
by a star and a black hole as the compact object). These systems usually spend 
long time in quiescence (${\rm L}_{\rm X}{\approx}10^{+33}$ erg/s) and
become in outburst (${\rm L}_{\rm X}{\approx}10^{+37}-10^{+38}$ erg/s) every few years.
During the outburst, spectral evolution occurs, but always beginning and 
ending in the low/hard state (described below).

\section{The relativistic ${\rm Fe}$ ${\rm K}_{\alpha}$ line}

Emission below 10\,keV can be due to a combination of thermal emission from the disk, and
non-thermal emission from a corona. It is difficult to study disks
through their continuum spectra, so the most pragmatic way to study the
disk is to use broad ${\rm Fe}$\,${\rm K}$ emission lines.
This line (as the reflection component that we will explain
later) arises from the fluorescence of the inner disk from an irradiating source
of high energy emission (George \& Fabian, 1991). Depending on the ionization
state of the disk this line is centered at 6.4-6.97\,keV. From the study of the
profiles of these lines, it can be inferred the spin of the black hole, which 
can be between zero for an Schwarzchild black hole (Fabian et~al., 1989) to a maximally
rotating regime in the case of a Kerr black hole (Laor, 1991). In the hard X-ray 
emission (${\rm E}{\geq}10$\,keV) a spectral hump usually appears, called the
reflection bump, with the same origin. Thus, both reflection and broad ${\rm Fe}$
${\rm K}_{\alpha}$ line should appear together during the observations. Nevertheless,
both effects are inhibited when a high ionization degree occurs, because the 
disk becomes a perfect reflector of the ${\gamma}$-rays (Ross et~al.1999).

Current observations of relativistic lines suggest that a
range of black hole spin parameters may be seen (for reviews, see
Miller 2007 and Nandra et al. 2007). It is therefore important to take
relativistic effects into account by fitting spectra with the
appropriate line and reflection models. Inner disks 
can get very close to the black hole (${\rm d}{\leq}6$\,${\rm R}_{\rm g}$) 
and this would imply a very concentrated source for the high energy emission, 
i.e., the high energy source has to be centrally compact.

\section{The states of the black holes and related behaviour}

In the most accepted picture for the black hole states (Tanaka \& Lewin, 1995)
the classification arises from the characterization of X-ray emission (usually a 
multicolor black body with an inner disk temperature of ${\rm T}<1$\,keV) and 
hard X-ray emission (${\rm E}{\geq}20$\,keV), the last best phenomenologically described
by a power-law. The high/soft state is characterized by strong disk emission ($>25\%$)
and an steep power-law (${\Gamma}>2.4$). The low/hard state has almost no disk emission
and the power-law dominates the spectrum ($>75\%$). During the Steep Power Law state
both disk and power-law emission are important, although ${\Gamma}$ is steep. 
During the evolution along the outburst some transitional states appear, called 
intermediate (Homan \& Belloni, 2005) with characteristics of both low/hard 
and high/soft states, depending on the case.

Grove et~al.(1998), on the basis of CGRO observations, proposed that the high/soft state
of black holes is associated with an unbroken powerlaw, usually understood as 
non-thermal origin of the high energy emission. On the other hand, the low/hard
state was associated with an spectral cut-off at 100\,keV (coinciding with the
kinetic temperature of the thermal distribution of the -inverse- Comptonizing
electrons), implying a thermal origin for the high energy emission. 

\section{The standard model}

Sunyaev \& Titarchuk (1980) proposed a thermal corona being the source of the
high energy emission from black-holes. Some years later, Esin et al.(1997) 
proposed the standard disk plus coronae model. In 
this model, the inner disk goes inwards while the outburst evolves from the 
low/hard to the high/soft states. An advection dominated accretion flow (ADAF)
corona fills the inner regions of the disk, providing the (inverse) Comptonizing 
electrons. The disk provides the photons to be Comptonized. This was considered 
to be the source of the high energy emission during the last 30 years.

However, the above mentioned model has several drawbacks. The first one is that
the standard model can not explain the QPOs (Quasi Periodic Oscillations) behaviour.
QPOs are the highest frequency oscillations in black holes (typically between 
140 and 450 Hz) which are commensurate with Keplerian orbital frequencies at the
ISCO (Inermost Stable Circular Orbit). X-rays QPOs appear in the low/hard state,
are maintained in the intermediate states and are quenched in the high/soft
state (see review by van der Klis, 2004). Also, there is an strong coupling 
between hard X-rays and radio emission, difficult to understand without 
taking into account hard X-ray emission from a jet. 
As can be seen in a recent review from Fender et~al.(2004), black
holes in the low/hard state show radio emission in the form of a steady jet with a
low value for the Lorentz factor (${\Gamma}<1.4$). During the intermediate state 
the radio emission is variable and ${\Gamma}$ is higher. In the Steep Power 
Law state the radio emission is optically
thin with a high value of ${\Gamma}>2$. In the high/soft state the radio emission 
is quenched. While the proposed scenario
in Fender et~al.(2004) would be consistent with the first point of the standard model
proposed by Esin et~al.(1997), the source of the high energy emission
in the low/hard state is still debated. The base of jets have been
proposed to be the source of the high energy emission Markoff et~al.(2001), 
Markoff et~al.(2003) and Markoff et~al.(2005). Nevertheless, the model of Markoff 
is only dedicated to hard X-ray domain ($E<100$\,keV).

\section{The alternative to the standard model}

Coppi (1999) developed the EQPAIR hybrid thermal/non-thermal (inverse) 
Comptonization model. This model can explain the excess (with respect a pure 
thermal model) of the high energy emission
already observed in black hole states during the high/soft state ($E>200$\,keV).
The physical mechanism of this model is that (inverse) Comptonization from a 
hybrid thermal and non-thermal distribution of particles (leptons) is the source
of the high energy emission. The most important consequences of this model are
both the disappearing of high energy cut-offs when the distribution of particles
is highly non-thermal and the presence of the anihilation line at 511\,keV.
This line results from the anihilation of relativistic $e^{+}e^{-}$ pairs in
mildly relativistic and non-thermal distributed plasmas. INTEGRAL and GLAST
are key missions in the detection of this line. Several attemps have been 
attempted to detect this line from microquasars but still there are not fiducial 
detections (Guessoum et al., 2006). We think that this line, if detected, 
would confirm the hypothesis of the source
of the high energy emission being highly concentrated and relativistic. These
conditions would be the typical of the base of a jet. Although this model seems
to be very promising it deals only with static coronae and, as shown in 
Beloborodov (1999), if particles of the plasma acquire high velocities 
($v/c{\geq}0.2$), the high energy radiation is highly anisotropic. 

\begin{figure}
\centerline{\psfig{file=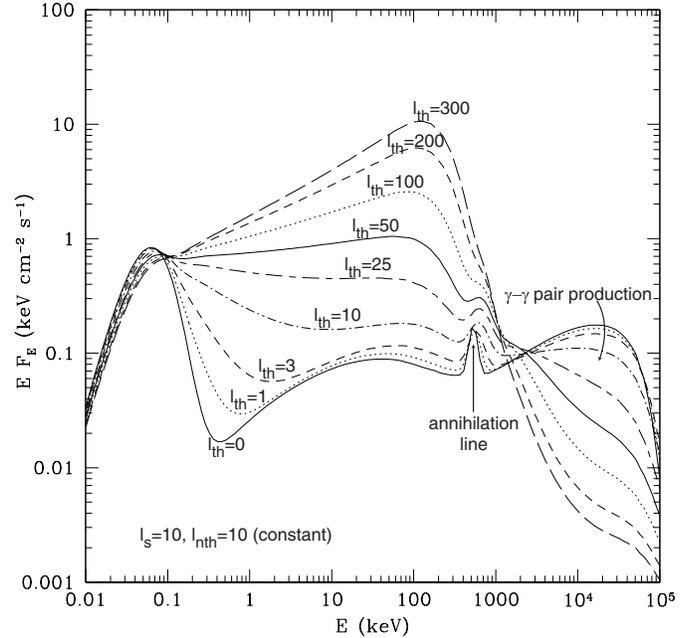,width=8.8cm,clip=} }
\caption{The transition from a non-thermal plasma (${\rm l}_{\rm th}=0$) to a 
thermally dominated plasma (${\rm l}_{\rm th}/{\rm l}_{\rm nth}=30$. The 
black-body spectrum has ${\rm T}_{\rm bb}=15$\,eV. Figure adapted from 
Coppi (1999). 
\label{image}}
\end{figure}

\begin{figure}
\centerline{\psfig{file=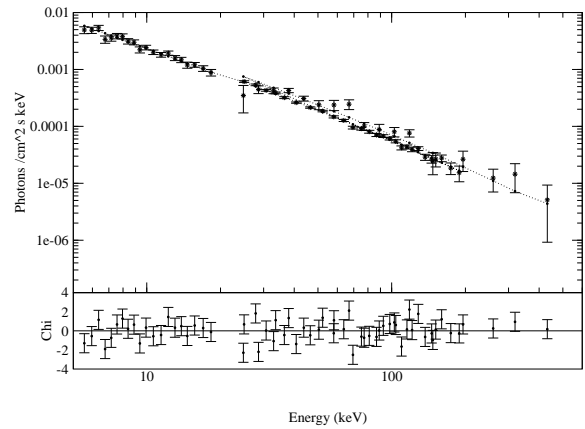,width=8.8cm,angle=270} }
\caption{Unfolded spectrum of GRO~J1655$-$40 in the low/hard state at the
beginning of the 2005 ourburst. Figure adapted from Caballero-Garc\'{\i}a
et~al.(2007).
\label{image}}
\end{figure}

\section{GRO~J1655$-$40: a particular case}
 
GRO~J1655$-$40 is a black hole transient discovered on 1994 July with the
CGRO satellite (Zhang et~al., 1994). It is a LMXB with a black hole of mass 
$m_{BH}=7.02{\pm}0.22M_{\odot}$ (Orosz \& Bailyn, 1997). The inclination
of the system is ${\approx}70^{\circ}$ (van der Hooft, 1998) although 
the inclination of the inner disk would be as high as $85^{\circ}$.

This system may harbor a maximal spinning black hole (${\rm a}{\geq}0.9$),
indicating an inner radius of  ${\rm r} \le 1.4$\,${\rm R}_{\rm g}$
(Miller et~al., 2005). Strong QPOs are detected (300-450\,Hz) (Strohmayer, 2001)
and extremely relativistic radio jets have been detected in radio 
(Hjellming \& Rupen, 1995). 

Tomsick et~al.(1999) did the former detection
of unbroken powerlaw emission up to 800\,keV during the low/hard state.
INTEGRAL observations during the 2005 outburst were done up to 500\,keV 
during the low/hard state (Caballero-Garc\'{\i}a et~al., 2007). In this work,
the low/hard state was modeled using both simple models like a power-law, 
and using EQPAIR. The lack of a break in fits with the simple power-law
implies that non-thermal Comptonization may dominate the hard X-ray
spectrum. Additional modeling with EQPAIR confirms that
non-thermal Comptonization is best able to explain the observed
spectrum. This work suggested thermal Comptonization processes
not being the main source for the high energy emission of 
GRO~J1655$-$40. However, more work is needed to disentangle the source 
of the high energy emission, and INTEGRAL observations are planned for 
a broad sample of black hole transients.


\end{document}